\documentclass[%
 reprint,
 amsmath,amssymb,
 aps,
]{revtex4-2}

\usepackage{graphicx}
\usepackage{dcolumn}
\usepackage{bm}
\usepackage[dvipsnames]{xcolor}
\usepackage{dsfont}
\usepackage{pgfplots}
\pgfplotsset{compat=1.15}
\usepackage{mathrsfs}
\usetikzlibrary{arrows}
\newcommand{\ket}[1]{| {#1} \rangle} 

\usepackage[utf8]{inputenc}
\usepackage[T1]{fontenc}
\usepackage[english,serbian]{babel}
\usepackage[OT1]{fontenc}
\usepackage{mathrsfs}
\usepackage{graphicx}
\usepackage{amssymb}
\usepackage{amsbsy}
\usepackage{latexsym}
\usepackage{mathtools}
\usepackage{titlesec}
\usepackage{amsmath}
\usepackage{color}
\usepackage[utf8]{inputenc}
\usepackage{lipsum}
\usepackage{fontenc}
\usepackage{bbold}
\usepackage{tikz-cd}

\usepackage{chngcntr}
\numberwithin{equation}{section}

\def\dj{d\kern-0.4em\char"16\kern-0.1em}

\def\diff{\textrm{d}}

\def\dj{d\kern-0.4em\char"16\kern-0.1em}
\def \Dj {\mbox{\raise0.3ex\hbox{-}\kern-0.4em D}}

\begin{document}
\definecolor{ududff}{rgb}{0.30196078431372547,0.30196078431372547,1.}
\definecolor{bcduew}{rgb}{0.7372549019607844,0.8313725490196079,0.9019607843137255}
\preprint{APS/123-QED}

\title{Testing the Braneworld Theory with Identical Particles}

\author{Ivana Stojiljkovi\'{c}, Du\v{s}an \Dj or\dj evi\'{c}, Aleksandra Go\v{c}anin,  and  Dragoljub Go\v{c}anin}
\affiliation{Faculty of Physics, University of Belgrade, Studentski Trg 12-16, 11000 Belgrade, Serbia}

\email{aleksandra.gocanin@ff.bg.ac.rs}

\selectlanguage{english}

\date{\today}

\begin{abstract}

Various attempts to go beyond the theory of General Relativity start from the assumption that spacetime is not a 4-dimensional but rather a higher-dimensional manifold. Among others, braneworld scenarios postulate that the spacetime we effectively observe is actually a 4-dimensional brane embedded in a higher-dimensional spacetime.
In general, braneworld models predict a departure from the Newton gravity law in the nonrelativistic regime. Based on this fact, we propose an experimental test that uses a pair of gravitationally interacting identical particles to determine the validity of certain braneworld models and provide numerical results that should be compared with experimental data. In particular, we consider the Randal-Sundrum braneworld model and study two cases of 5-dimensional gravity theories: the Einstein-Hilbert gravity with the negative cosmological constant and the Einstein-Gauss-Bonnet (nearly-Chern-Simons) gravity. 

\end{abstract}

\maketitle

\selectlanguage{english}

\section{Introduction}

Newton's law of gravity has been experimentally verified a long time ago \cite{Cavendish} and since then has been repeatedly tested using various experimental setups, both in the lab and in cosmological observations. However, we know that this law is only an approximation, as a more fundamental theory is general relativity (GR). Although its predictions have been corroborated in many different ways to a high level of precision \cite{GREXP, LIGO}, we are aware that GR does not amount to a full story about spacetime and gravity. The main symptom of the theory's incompleteness is the appearance of singularities, both in the centre of a black hole and at the beginning of the universe. In order to solve this problem, many attempts to quantize gravity have been made \cite{Carlip}. The most naive procedure is based on using a linear approximation of Einstein's field equations, which leads to a non-renormalizable theory, and, therefore, cannot be trusted to arbitrary high energies. Therefore, it seems that some radical change in our basic assumptions has to be made when dealing with the quantum theory of gravity. Some approaches, among which string theory is perhaps the most notable one, start from a theory defined in a number of spacetime dimensions greater than four. Effective, 4-dimensional physics is then obtained from the Kalutza-Klein compactification \cite{Kaluza, Klein}, where one assumes that additional spatial dimensions are small enough and thus experimentally inaccessible at current energies. 

For phenomenological reasons, an alternative procedure of eliminating extra dimensions was proposed that does not assume the existence of small extra dimensions \cite{Randall:1999ee, Randall:1999vf, Arkani-Hamed:1998jmv}. Those models are known under the name of braneworld models and assume that our world is a 3-brane, embedded in a higher-dimensional spacetime. Matter fields are usually confined on this brane, while gravity is a priori free to propagate in all dimensions. Naturally, those models lead to some conclusions that differ from the standard gravity theories. In this paper, we will propose a way to experimentally test the predictions of some braneworld scenarios.

\section{Braneworld scenarios}
For simplicity, we will mostly focus on the so-called Randal-Sundrum (RS) II model (see \cite{Bilic:2015uol} for a review). We start from a 5D Einstein gravity theory with the negative cosmological constant, where the action is defined as 
\begin{equation}
    \frac{1}{16\pi G}\int_{\mathcal{M}_5} \diff^5x \; \sqrt{-g}\left(R-2\Lambda    \right )+\frac{1}{8\pi G}\int_{\partial \mathcal{M}_5}\diff ^4x \sqrt{-h}K.
\end{equation}
The Gibbons-Hawking-York term has to be included in the presence of a boundary so that the variation principle is satisfied. We then insert a brane $Q$ of constant tension $T$, usually defined as a hyper-surface for which one coordinate in a preferred coordinate system is constant. The relevant term that we add to the action is 
\begin{equation}
    T\int_{Q}\diff ^4x \;\sqrt{-h}.
\end{equation}
Away from the brane, the anti-de Sitter ($\mathrm{AdS}$) spacetime solves equations of motion. However,  appropriate junction conditions have to be imposed that patch together solutions on both sides of the brane. Also, more complicated matter fields that are localised on the brane could be added.

In order to derive Newton's law of gravity from underlying relativistic theory, one has to use the action and equations of motion, and therefore for different theories, we can get different results.
Generically, we write the potential on the brane in the form
\begin{equation}
    V(r)=-\frac{Gm}{r}(1+\Delta(r)).
\end{equation}
In the case of RS-II model, for distances $r\gg \frac{1}{k}$ ($k$ is the inverse of the bulk AdS radius $l_{\text{AdS}}$) we have \cite{Callin:2004py, Bronnikov:2006jy}
\begin{equation}
    \Delta(r)=\frac{2}{3k^2r^2}-\frac{4\ln kr}{k^4r^4}+\frac{(16-12\ln 2)}{3k^4r^4}+\dots
\end{equation}
while for small distances ($r\ll \frac{1}{k}$),
\begin{equation}
    \Delta(r)=\frac{4}{3\pi kr}-\frac{1}{3}-\frac{kr}{2\pi}\ln kr+0.089237810 kr+\dots
\end{equation}
One can also find an approximate potential interpolating between those two extremes. It is given by \cite{Deruelle:2003tz}
\begin{align}\label{RSII}
    &\Delta(r)=\frac{4}{3\pi}\Big ( \frac{kr\cos kr-\sin kr}{k^2r^2}\int_{+\infty}^{kr}\frac{\cos t}{t}\diff t \nonumber\\ 
    &+\frac{\cos kr+kr\sin kr}{k^2r^2}\int_{+\infty}^{kr}\frac{\sin t}{t}\diff t+\frac{\pi }{2k^2r^2}
    \Big).
\end{align}
There are other braneworld models that one could also study. For example, in \cite{Parvizi:2015uda}, modified Newton potential was derived, with the following large distances behaviour 
\begin{equation}
    \Delta(r)=-\frac{e^{-2\sqrt{2}kr}}{4(\sqrt{2}-1)kr},
\end{equation}
where $k\gg\frac{1}{r}$ is the inverse length parameter. There are also models that start from a different number of spacetime dimensions or that involve more than one brane \cite{Bronnikov:2006jy}. Attempts to experimentally verify those corrections have been made in the context of classical physics \cite{Azam:2007ba}.

It is important to note that results concerning the modification of Newton's potential follow from the tree-level computations and therefore do not contain nonzero powers of $\hbar$. This is in contrast with the one-loop calculations of the gravity propagator that yields a similar result (though using the ideas from AdS/CFT, those two can be seen on equal footing \cite{Duff:2000mt}). Namely, if we were to quantize GR (despite it being non-renormalizable), we could obtain modifications of Newton's potential at the one-loop level, but those would be suppressed by powers of $\hbar$. The potentials that we analyse in this paper are not of this type. 
This means that we don't have to claim any results from quantum gravity in order to talk about those modified potentials; we only assume the existence of additional dimensions. Also, corrections of the form $\frac{1}{r^3}$ in Newton's potential are well-known in GR (they are responsible for the precession of Mercury orbit). However, they are not in any way connected to the corrections we are dealing with here, which are nonrelativistic. 

\section{Two identical particles}

Let us consider two identical spin-$\frac{1}{2}$ particles that have zero charge with respect to any internal symmetry. This means that the only way those two particles can interact is via gravitational force. For small masses (energies), classical gravity is well-approximated by Newton's theory, which allows us to assume the following form of the two-particle Hamiltonian,
\begin{equation}\label{ham}
    H=\frac{\boldsymbol{P}^2}{4m}+\frac{\boldsymbol{p}^2}{m}-\frac{Gm^2}{r},
\end{equation}
where $\boldsymbol{P}$ is total momentum of the system and $\boldsymbol{p}$ is relative momentum. 

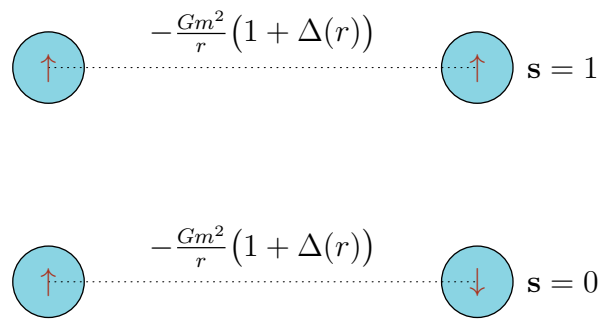
\begin{figure}[h!]
    \centering
     \begin{tikzpicture}[scale=0.95,line cap=round,line join=round,x=1.0cm,y=1.0cm]
\clip(-2.371785018826553,-3.4718228958881) rectangle (12.396448434326285,2.404158982779382);
\filldraw [fill=Turquoise!50,line width=0.5pt] (-1.5,1.) circle (0.5cm);
\filldraw [fill=Turquoise!50,line width=0.5pt] (4.5,1.) circle (0.5cm);
\filldraw [fill=Turquoise!50,line width=0.5pt] (-1.5,-2.) circle (0.5cm);
\filldraw [fill=Turquoise!50,line width=0.5pt] (4.5,-2.) circle (0.5cm);
\draw[dotted] (-1.5,1.)--(4.5,1.);
\draw[dotted] (-1.5,-2.)--(4.5,-2.);
\draw (1.5,1.5) node {{\large{$-\frac{Gm^2}{r}\big(1+\Delta(r)\big)$} }};
\draw (1.5,-1.5) node {{\large{$-\frac{Gm^2}{r}\big(1+\Delta(r)\big)$} }};
\draw (-1.5,1.)  node {\color{Mahogany}{\large{$\uparrow$}}};
\draw (4.5,1.)  node {\color{Mahogany}{\large{$\uparrow$}}};
\draw (-1.5,-2.)  node {\color{Mahogany}{\large{$\uparrow$}}};
\draw (4.5,-2.)  node {\color{Mahogany}{\large{$\downarrow$}}};
\draw (5.7,1.) node {{\large{$\mathbf{s}=1$}}};
\draw (5.7,-2.) node {{\large{$\mathbf{s}=0$}}};
\end{tikzpicture}
    \caption{Two identical spin-$\frac{1}{2}$ fermions with zero charges, interacting through a modified Newton's potential. Depending on the spin polarisation, the total energy of the system is different.}
    \label{skicaidenticnih}
\end{figure}

Eigenstates of the hamiltonian (\ref{ham}) are of the form 
\begin{equation}
    \ket{\psi}=\frac{e^{i\Vec{K}\cdot \Vec{R}}}{\sqrt{V}}R_{nl}(r)Y_{l}^{\;m}(\theta,\varphi)\otimes \ket{\chi}.
\end{equation}
We assume that the spatial volume $V$ of the box in which we constrain our quantum system is much larger than relevant scales of bound states of Hamiltonian (\ref{ham}) so that we can neglect the influence of this box on the spectrum. In case we place the system in some external (non-constant) potential, we demand that the relevant scales o variations in the potential are large compared to any scales in (\ref{ham}) so that we can again assume that we can restrict ourselves to the case of Hamiltonian (\ref{ham}).
The interchange of two particles corresponds to $\Vec{r}\rightarrow -\Vec{r}$, which induces the change $Y_{l}^{\;m}(\theta,\varphi)\rightarrow (-1)^l Y_{l}^{\;m}(\theta,\varphi)$. Preparing the spin state of the system in an appropriate manner (singlet or triplet), we are able to control the parity of relative angular momentum of eigenstates, as the total state of the two particles has to be antisymmetric. In the case of $\frac{1}{r}$ potential, states with different values of $l$ have the same energy as long as the principal quantum number is the same. This is related to the $SO(4)$ symmetry of the Hamiltonian (\ref{ham}). However, in the case of braneworld models, this Hamiltonian is corrected using previously described potentials. New Hamiltonian has only rotational $SO(3)$ symmetry, and therefore states with different angular momentum have different energy. We want to determine the eigenenergies of the braneworld Hamiltonians resulting from various braneworld scenarios. First, let us consider the RSII model with Einstein-Hilbert action, where correction to Newton's potential is given by (\ref{RSII}).  

\begin{figure}
    \centering
\includegraphics[width=0.45\textwidth]{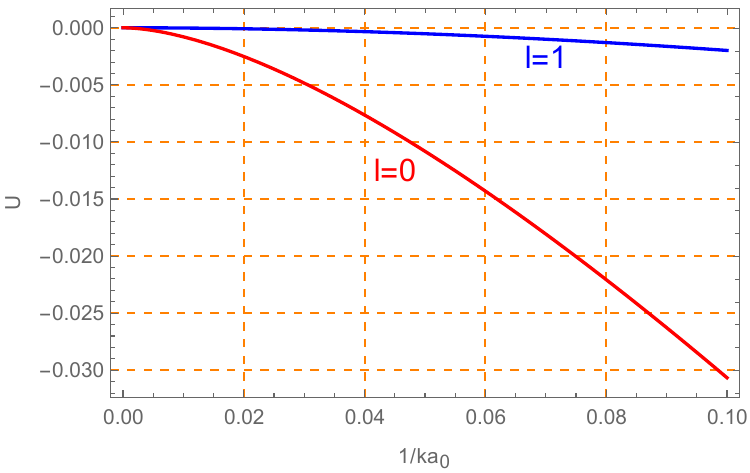}
    \caption{Corrections to the energies of the first excited level for $l=0$ (red line) and $l=1$ (blue line); color online.}
    \label{obagrafika}
\end{figure}

There are two regimes that we will consider. The first one is when the perturbation theory is applicable. Eigenenergies of the unperturbed Hamiltonian (\ref{ham}) are well-known, as it is the same (up to a numerical factor) as the Hamiltonian of the hydrogen atom. This means that the energies (disregarding the centre of mass energy) are given by 
\begin{equation}
  E_n=  -\frac{a^2\hbar^2}{mn^2},
\end{equation}
where $a=\frac{G m^2}{\hbar^2}$. 
We now consider correction to the potential $-\frac{Gm^2(1+\Delta(r))}{r}$ as a perturbation and use the first-order perturbation calculus to obtain corrections to the energies of the $n=2$ level. We will see that the perturbation calculus is valid as long as $1\ll ka_0$, where $a_0=\frac{2\hbar^2}{Gm^3}$. As the perturbation also has $SO(3)$ symmetry, we only need to calculate diagonal matrix elements and obtain corrections to the energy levels. In Figure \ref{obagrafika}, we present those corrections as a function of the dimensionless parameter $\frac{1}{ka_0}$.
To be more precise, we plot dimensionless quantity 
\begin{equation}
    U=8\frac{a_0}{k^2}\int_0^{+\infty} \diff x\; |\psi(x)|^2(-x)\Delta(x),
\end{equation}
such that first-order corrections to energy levels are given by $E'=\frac{Gm^2}{8a_0}U$.
It is important to note that our system is nonrelativistic so that we can use Newtonian approximation. For this to be true, we must have $p\sim\frac{\hbar}{a_0}\ll mc$. 
In the case when the perturbation theory is not applicable, we can make the following observation. In this regime, $a_0$ is at least of the same order as $1/k$, or possibly larger. Wave functions peak around $a_0$, and this means that we can use the form of the potential for $r\ll \frac{1}{k}$. It turns out that potential $-\frac{A}{r}-\frac{B}{r^2}$ corresponds to a Hamiltonian that we can exactly diagonalize. It can be shown that for $\frac{1}{ka_0}>\frac{3\pi}{32}\approx 0.3$, the particles would merge, and our analysis based on quantum mechanics and Newton's gravity would be inappropriate \cite{LandauL}.
\begin{figure}[h!]
    \centering
    \includegraphics[width=0.45\textwidth]{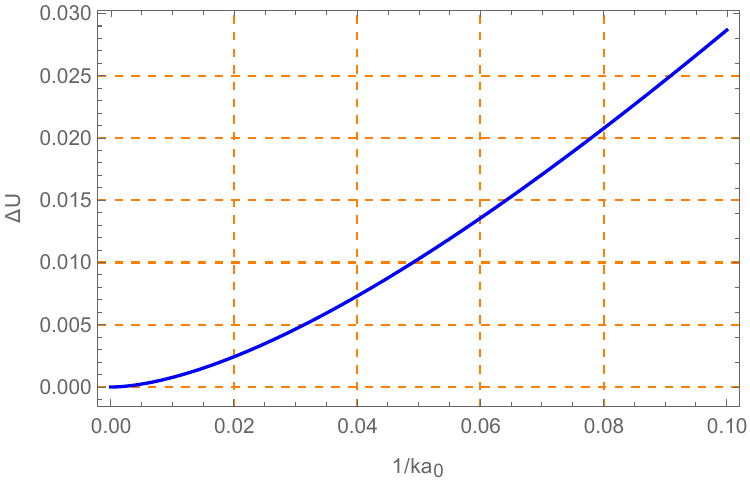}
    \caption{Difference of energies between $l=0$ and $l=1$ state for the first excited level.}
    \label{figCS}
\end{figure}

\subsection{Chern-Simons gravity}
So far, we assumed that the 5-dimensional theory is well-described by Einstein's gravity. In five dimensions, one can also consider Chern-Simons (CS) gravity, defined in the metric formulation as 
\begin{align}\label{CSaction}
S_{CS}=\frac{1}{16\pi G}&\int\diff^{5}x\sqrt{-g}\Big[R-2\Lambda\\
+&\frac{1}{4k^2}\left(R^{2}-4R^{\mu\nu}R_{\nu\mu}+R^{\mu\nu\rho\sigma}R_{\rho\sigma\mu\nu}\right)\Big]\nonumber .\end{align}
This theory has an enlarged $SO(4,2)$ symmetry group \cite{CS_book}. CS theory has non-vanishing torsion, but here we restrict to the case of torsion-less geometries, as they are much better understood. 
More generally, the CS action (\ref{CSaction}) can be generalized by substituting the constant parameter $\frac{1}{4k^2}$ with some general parameter $\frac{\alpha}{4k^2}$, thus obtaining the Einstein-Gauss-Bonnet theory. One can calculate the modification of Newton's potential for this generalized gravity theory \cite{Deruelle:2003tz} and obtain that, for the CS case, there are no corrections. Moreover, if we take $\alpha$ close to the CS value, we get corrections that are small enough so that perturbation theory can be used for all values of $\frac{1}{ka_0}$. 

The correction to the potential takes the form \cite{Deruelle:2003tz}
\begin{align}
\nonumber
&\Delta_{\alpha}(r)=\frac{4 (1-\alpha)}{3\pi 
(1+\alpha)}\Bigg(\frac{(\beta x \cos (\beta x)-\sin (\beta x)) \int_{+\infty }^{\beta x} \frac{\cos (t)}{t} \,
   \diff t}{x^2}\\\nonumber
   &+\frac{(\cos (\beta x)+\beta x \sin (\beta x)) \int_{+\infty }^{\beta x} \frac{\sin (t)}{t} \, \diff t}{x^2}+\frac{\pi }{2
   x^2}-
   \frac{\beta}{x}+\beta^2\gamma   \\
   &\times\big(\sin (\beta
   \gamma  x)\int_{+\infty }^{\beta \gamma  x} \frac{\cos (t)}{t} \, \diff t -\cos (\beta \gamma  x) \int_{+\infty }^{\beta \gamma  x} \frac{\sin (t)}{t} \, \diff t\big)\Bigg),
\end{align}
where $x=kr$, and $\beta$ and $\gamma$ are numerical constants defined in \cite{Deruelle:2003tz}; for $\alpha=0.95$ they are given as $\beta\approx3.86443$ and $\gamma\approx0.637448$. Again, we are using approximate potential, with approximate numerical values of the parameters, as our goal is to demonstrate the possibility of detection of extra dimensions and provide some rough theoretical data that could be made more precise with more advanced numerical techniques. 
In Figure \ref{figCS}, we present corrections to the energy level $n=2$, $l=0$, and $n=2$, $l=1$ in the case $\alpha=0.95$. Note that the form of the corrections is expected. The energy is shifted more for greater values of $\frac{1}{ka_0}$, but the rate of this shift is decreasing. Also, for large $\frac{1}{ka_0}$, the energy difference between the two levels decreases. This can be concluded from the asymptotic form of the gravitational potential.

\begin{figure}[h!]
    \centering
    \includegraphics[width=0.45\textwidth]{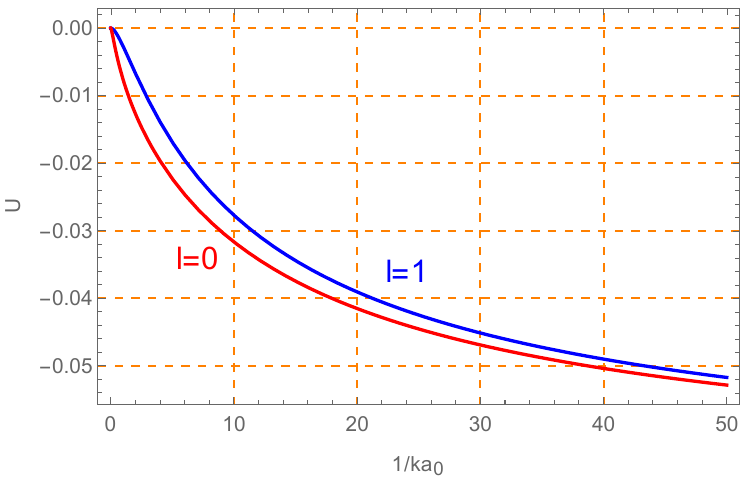}
    \caption{Corrections to the energies of the first excited level for $l=0$ (red line) and $l=1$ (blue line) in nearly-CS case; color online.}
    \label{figCS}
\end{figure}

\begin{figure}[h!]
    \centering
    \includegraphics[width=0.45\textwidth]{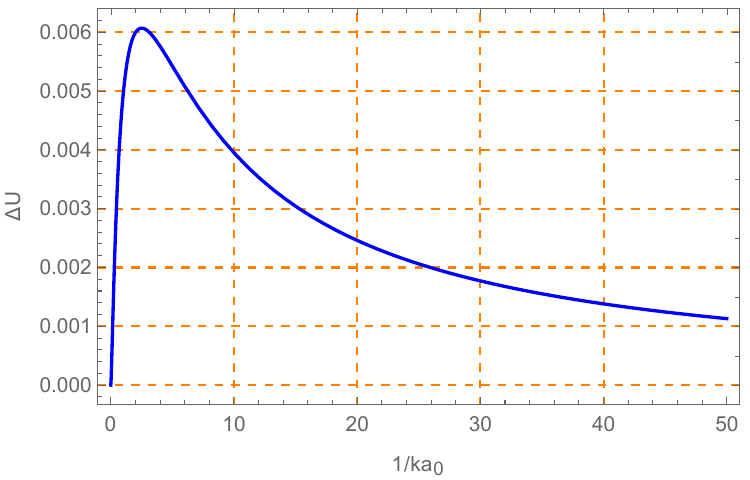}
    \caption{Difference of energies between $l=0$ and $l=1$ state for the first excited level in the nearly-CS case.}
    \label{figCS1}
\end{figure}

Finally, we can draw the difference between energies of $l=0$ and $l=1$ level, obtaining Figure \ref{figCS1}. Due to the approximate nature of our constants, previous graphs may not give the best numerical values for large $\frac{1}{ka_0}$, but the form of the graph should be correct.

\section{Testing the braneworld hypothesis}

\begin{figure}[h!]
    \centering
\includegraphics[width=0.5\textwidth]{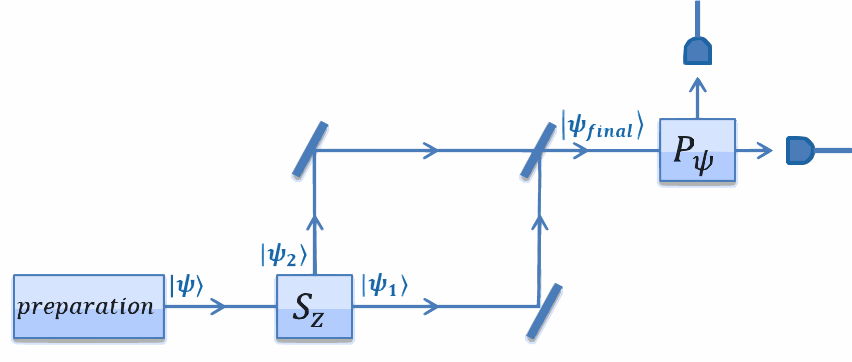}
    \caption{Proposed experimental setup. Interferometry is used to determine the phase shift between two states and therefore determine the existence of braneworld models.}
    \label{figexp}
\end{figure}

Let us now propose a way to empirically verify whether the consequences of the braneworld hypothesis are valid or not. The procedure is essentially based on the fact that the beyond-Newtonian gravitational potential (of the kind we considered above) lifts the orbital degeneracy of the energy levels for a pair of gravitationally interacting particles. Consider, for example, the following two states of a pair of identical spin-$\tfrac{1}{2}$ particles,
\begin{align}
\vert\Psi_{1}\rangle&=\frac{1}{\sqrt{2}}\vert 2, 0, 0\rangle\otimes(\vert\uparrow\downarrow\rangle-\vert\downarrow\uparrow\rangle),\\
\vert\Psi_{2}\rangle&=\vert 2, 1, 0\rangle\otimes\vert\uparrow\uparrow\rangle.
\end{align}
Both states are anti-symmetric as a whole, the particles being fermions. The first state has the anti-symmetric singlet state in the spin sector, while the orbital part is symmetric ($l=0$). On the other hand, the second state has a symmetric spin part and an anti-symmetric orbital part (note that we could also choose some other spin state from the symmetric triplet and also some other value for the magnetic quantum number, namely $\pm 1$). The braneworld potential (\ref{RSII}) implies that states with different orbital quantum numbers $l$ have a different energy, which is in contrast with the Newtonian case. In principle, we may also consider a superposition of the above two states,
\begin{equation}\label{Psi}
\vert\Psi\rangle=\frac{1}{\sqrt{2}}(\vert\Psi_{1}\rangle+\vert\Psi_{2}\rangle),
\end{equation}
and use the standard Mach-Zehnder (MZ) interferometer \cite{MZ} with the first beam splitter selecting the states by the total spin projection $S_{z}$. If the energies of the states $\vert\Psi_{1}\rangle$ and $\vert\Psi_{2}\rangle$ were the same, as in the Newtonian case, the final state after passing through the MZ interferometer would be the same, namely, $\vert\Psi\rangle$.
If, on the other hand, the energies of the two states were different, as predicted by the braneworld model, the final state of the two particles would be
\begin{equation}
\vert\Psi_{\text{final}}\rangle=\frac{1}{\sqrt{2}}(e^{-i\phi}\vert\Psi_{1}\rangle+\vert\Psi_{2}\rangle),
\end{equation}
where the relative phase is determined by the energy difference,
\begin{equation}
\phi=(E_{n=2,l=1}-E_{n=2,l=0})t/\hbar. 
\end{equation}
The second beam splitter in the MZ scheme should post-select the state $\vert\Psi \rangle$. If $\phi$ is non-zero, i.e. if the braneworld corrections exist, the statistics of detectors' clicks would be modified to $\cos^2 \frac{\phi}{2}$ and $\sin^2 \frac{\phi}{2}$. In this way, we can experimentally obtain the value of the energy gap if it exists. Therefore, if one could put our particles in the superposition state $\vert\Psi\rangle$, simple MZ interferometry could be used to refute (or support) the braneworld scenario. If the outcome of the experiment were positive, in support of the braneworld, one could estimate the initially undetermined bulk parameter $\frac{1}{k}=l_{\text{AdS}}$. In Figure \ref{figexp}, we give a scheme of the experimental proposal.  

\section{Conclusion}

In this work, we explored the possibility of testing the braneworld scenario in laboratory conditions. We have demonstrated that the braneworld RS-II model predicts a modification of the energy spectrum of gravitational bound states by lifting the characteristic orbital degeneracy associated with the Newtonian case. For a pair of identical fermions, the energy spectrum depends on whether the system is in the singlet or triplet spin state. From a practical point of view, the crucial step would be to prepare the particles in the superposition state (\ref{Psi}). Due to the fermionic symmetry constraint, this might be achieved by solely manipulating the spins of the two particles, which is the main reason for working with a pair of identical fermions. We studied two cases of 5-dimensional theories of gravity - Einstein's gravity and nearly-Chern-Simons gravity - and presented the numerical results concerning energy splitting, which appears solely due to the braneworld hypothesis. To test the existence of the splitting, one could, in principle, set up a Mach-Zehnder interferometer, as explained in the text. We believe that bounded gravity states, in combination with quantum mechanics, could be the most straightforward way to test the braneworld models. 
Controlling the masses of the particles (thus controlling the parameter $a_0$), one can obtain a graph of energy difference and compare it to one of the figures from the previous section. In this way, one could discriminate between different models of 5-dimensional gravity. Another proposal for testing the RS model can be found in \cite{Azam:2007ba}.   

A possible extension of our work would be to make a comparison to a similar experimental set-up \cite{Bose:2017nin, Marletto:2017kzi}, discussing the entanglement between quantum particles induced by gravity \cite{Yant:2023smr, Kanno:2021gpt,vandeKamp:2020rqh, Schut:2023hsy, Hanif:2023fto}. In particular, one could use potentials analysed in this paper to see the differences induced by the corrections coming from the RS-II model. In the case of EH gravity, and using the limit of small or large distance, this was done recently in \cite{Elahi:2023ozf, Feng:2023krm}. It would be interesting to analyse other experimental proposals and to see the effects coming from different gravity models in the 5-dimensional bulk. 
Effects analysed in this work could also be relevant for cosmology \cite{Nielsen:2019izz}.

\section{Acknowledgement}
The authors express their gratitude to \v{C}aslav Brukner for his hospitality and productive discussions during their visit to the Institute for Quantum Optics and Quantum Information (IQOQI) Vienna. The authors also thank A.C. de la Hamette and V. S. Kabel for useful   discussion on the content of this paper. This paper is produced as a result of the BBQUANT project, with active participation from A.G. and D.G. This project was conducted within the Serbian Science and Diaspora Collaboration Program: Knowledge Exchange Vouchers, supported by the Science Fund of the Republic of Serbia.
Work of I.S., D.\Dj., A.G. and D.G. is supported by the funding provided by the Faculty of Physics, University of Belgrade, through grant number 451-03-47/2023-01/200162
by the Ministry of Science, Technological Development
and Innovations of the Republic of Serbia.

\end{document}